\documentclass[twocolumn,preprintnumbers,amsmath,amssymb,prb]{revtex4}

\usepackage{amsmath}
\newcommand{\lb}{{<}}
\newcommand{\rb}{{>}}
\newcommand{\srm}{{\rm S}}
\newcommand{\vrm}{{\rm V}}

\begin{document}

\title {Comment on \lq\lq On Mach's critique of Newton and Copernicus \rq\rq}
\author{A. Bhadra} 
\email{aru_bhadra@yahoo.com}
\affiliation{High Energy and Cosmic Ray Research Centre, University of North Bengal, Siliguri, WB 734013 
India}
\author {S. C. Das}
\affiliation{Administrative Block, University of North Bengal, Siliguri, WB 734013 
India}

\begin{abstract}
Hartman and Nissim-Sabat have argued that Mach's idea of the relativity of rotational motion suffers from internal inconsistencies and leads to a contradiction that there cannot be a stationary bucket in a rotating universe. They also claimed that non-inertial electromagnetic and stellar aberration observations can distinguish between a rotating and a stationary universe, whereas according to Mach there cannot be any observable way to distinguish these two cases. We contest these objections. 
\end{abstract}

\maketitle

\section{Introduction}
There is a long-standing debate on whether space is absolute or relative. The question has profound physical significance because all physical processes take part in space and time. Newton favored the absolute nature of space (and time) and cited his famous rotating bucket experiment to support his view.\cite{ab:1} However, Huygens, Leibniz, Berkeley, Einstein were among the notable critics of Newton's absolute space.\cite{ab:2}$^{,}$ \cite{ab:3} But the most serious conceptual objections to the notion of absolute space came from Mach who advocated the relativity of all motion including rotational motion. \cite{ab:4}$^{,}$ \cite{ab:3}

Hartman and Nissim-Sabat\cite{ab:5} (HN) have recently argued that Mach's idea of the relativity of rotational motion suffers from internal inconsistencies and leads to a contradiction that there cannot be a stationary bucket in a rotating universe. HN limited their discussion of Mach's ideas to classical physics, claiming that Mach should have addressed the issues that they raised. They further claimed that electromagnetic and stellar aberration observations could distinguish between a rotating universe and a fixed one, whereas according to Mach there cannot be any observable way to differentiate these two cases. Many other scientists including Plank, Einstein, and Feynman \cite{ab:3}$^{,}$ \cite{ab:6} have criticized the notion of the relativity of rotation on different grounds, but Ref.~\onlinecite{ab:5} is very comprehensive and deserves special attention.

In this paper we consider the arguments against Mach's principle raised in Ref.~\onlinecite{ab:5}. In Sec.~\ref{sec2} we briefly describe the famous rotating bucket experiment and its Newtonian and Machian interpretations. We also discuss briefly the Machian philosophy of relativity. In Sec.~\ref{sec3} we examine whether any contradiction occurs if a bucket is kept fixed in a rotating universe. After examining critically the physical effects cited in Ref.~\onlinecite{ab:5} to detect an absolute state of motion, we give our arguments against such proposals in Sec.~\ref{sec4}. The equivalence of heliocentric and geocentric systems is discussed in Sec.~\ref{sec5}. We summarize our discussion in Sec.~\ref{sec6}.

\section{Bucket experiment, absolute space and principle of relativity of all motion}\label{sec2}

A frame of reference is a prerequisite to representing an object spatially and temporally. Newton postulated the existence of an absolute space to describe the absolute motion of an object, which implies that there is an absolute frame of reference rigidly attached to absolute space that provides a unique way of identifying spatial locations through time. A major objection against absolute space was that it couldn't be distinguished from all the other inertial frames by any observational properties. Acceleration in rotational motion, as considered in the bucket experiment, is supposed to provide a way to differentiate the absolute frame from other frames.

The bucket experiment is simple. A bucket half-filled with water is suspended from a fixed point by a twisted rope and the bucket is let go. The bucket begins to rotate as a result of the unwinding of the rope. Initially the water in the bucket does not follow the motion of the bucket but remains stationary and its surface remains flat. Gradually, the motion of the bucket is communicated to the water and the surface of the water becomes concave. 

Why does the surface of the water become curved? Is it because of the rotation of the water? But what does rotation mean here (with respect to what)? Certainly it is not relative to the bucket as both the bucket and water are spinning with the same angular velocity. To avoid local influences Newton considered a thought experiment where the experiment is conducted in empty space where there is nothing to measure the rotation. Most physicists were believed that the outcome of the experiment would be the same, namely the water surface would become concave even in such a hypothetical experiment. Newton thus concluded that there had to be an absolute frame relative to which the rotation can be measured.

Mach disagreed with Newton's interpretation. He assumed that all motion, including rotational motion, is the relative motion of material bodies. To Mach, the inertia of a particle is due to some (unspecified) interaction of the particle with all the other particles of the universe and the local measures of non-acceleration are determined by an average of the motions of all the masses in the universe.\cite{ab:7} For Mach centrifugal forces are gravitational, that is, they are produced by the interaction of one particle with another. Mach argued that the results of the bucket experiment could equally be explained by considering the rotation of the water relative to all the matter in the universe, instead of the rotation of the water relative to absolute space. In the absence of all other particles in the universe the result of the experiment would be different: the surface of the water would never become concave. Because this experiment cannot be performed, it is impossible to determine whether Mach or Newton is right. 

The current understanding of the relativity of rotation, which is usually named after Mach, is also due to many other physicists who contributed interesting ideas, both in support and opposition to Mach's position. Mach's ideas on inertia were already contained in the writings of Leibniz and Berkeley. Hofmann\cite{ab:8} explained the origin of inertial forces by stating that just as an object experiences centrifugal forces when it rotates with respect to the rest of the universe, all other particles in the universe also should experience centrifugal forces because we could equally consider the rotating object as fixed, while the rest of the universe is rotating about it. However, the state of motion of all the particles of the universe will be dominantly governed by the overwhelming mass of the rest of the universe, and hence there will be no measurable centrifugal phenomena in the frame of rest of the universe.  Einstein first demonstrated by an ingenious gedanken experiment that a rotating mass shell exerts inertial forces on test particles near the center of the shell.\cite{ab:9} It is now generally believed that the dragging of the inertial frames by rotating masses is a Machian effect (however, see Ref.~\onlinecite{ab:10}). Lense and Thirring calculated the orbital precession due to rotation of the central source.\cite{ab:11} Although the Lense-Thirring precision of planetary orbits is small and not measurable with present technology, the effect may be detected by artificial satellites orbiting the earth. It is one of the major goals of the Gravity Probe B experiment.\cite{ab:12} Machian ideas have also motivated a new approach to quantizing gravity.\cite{ab:13} 

\section{Fixed bucket in the rotating universe}\label{sec3}

HN claimed in Secs.~IIA and IIF of Ref.~\onlinecite{ab:5} that a fixed bucket in a rotating universe is not consistent with Mach's ideas on the relativity of rotational motion. They considered a system consisting of a bucket, a star (representing the universe) and a pendulum that is mounted above the bucket and is swinging with the same frequency as the bucket's rotation. The bucket has a V painted on the bottom, the star has a bright spot marked by S${>}$, and the bob of the pendulum is marked by $\Delta$. In the initial orientation the bob is between the star and the bucket as the pendulum is fully extended. They demonstrated that the relative positions and orientations of the bucket, star, and the bob are different when the bucket rotates (BR) in a stationary universe, and when a stationary bucket is in a rotating universe (UR) half a period later. In the UR case, the bucket will never be in between the star and the pendulum bob. According to Ref.~\onlinecite{ab:5} the position and orientations initially and half a period later in the BR and UR cases are
\begin{eqnarray}
\begin{array}{llll}
\mbox{Initial:} & \srm\rb & \Delta & \vrm \\
\mbox{BR:} & \srm\rb & \Lambda & \Delta\\
\mbox{UR:} & \vrm &\Delta & \lb\srm
\end{array}
\end{eqnarray}
It appears from the configurations that the observer is rigidly attached with the universe, not with the bucket, and the observer is facing the pendulum bob. Next HN assumed that the rotating universe induces the motion of the pendulum. In such a situation the configuration at half a period later would be 
\begin{eqnarray}
\begin{array}{llll}
\mbox{UR:} & \Delta & \vrm & \lb \srm,
\end{array}
\end{eqnarray}
which is still distinguishable from the BR case. HN noted that the BR configuration can be duplicated if the rotation of the universe induces a rotation of the pendulum bob as well. They argued that if a rotation is induced in the bob, a rotation must also be induced in the bucket. These observations led HN to conclude that Mach's ideas imply that there cannot be a fixed bucket in a rotating universe. 

Without going into any analysis we can say that this conclusion cannot be true. The reason is that if the bucket rotates with the universe, then the relative positions of the bucket and the distant fixed star would remain the same, which is different from the BR scenario. To examine the other cases let us consider a parallel but simpler case in linear unaccelerated motion. Let three objects, A, B, and C, initially located apart on a straight line with B in between A and C be at rest with respect to each other. Now let A start moving with a constant velocity toward C, B being at rest. Because non-accelerating motions are relative in nature, the motion of A toward the fixed C is the same as the motion of C toward the fixed A. But in the first case (moving A), A will cross B after a certain time whereas in the second case (moving C) such an event will never occur. So the two motions are not identical. The reason is that B is taken as rest with respect to C in the first case but not in the second case. The above example implies that while considering relative motion we cannot invoke an independent third frame of reference. 

If we take B to be rigidly fixed with A (or C) so that they effectively become one system, there would be no difference whether we take A moving and C at rest or vice versa. Similarly it is expected that BR and UR with induced pendulum motion would be indistinguishable. However, HN found that they are not because they did not consider the position of the observer in the UR case. It is natural that anything within the universe will move with it. So both the observer and the pendulum will rotate as the universe rotates (not because of any frame dragging effect). But in the UR scenario, the bucket is kept fixed (by applying external forces just like as is needed to rotate the bucket in a static universe) and is not allowed to move with the universe. If we consider such facts, the orientations of the objects with respect to the observer half a period later from the initial position would be the same as the BR case. 

HN also pointed out that the principle of the relativity of rotation requires that in the UR case, S$\rb$ would become $\lb$S rather than remaining S$\rb$ half a period later from the initial configuration. Such a change means that S rotates as it revolves with respect to the bucket. We differ with this assertion. Rotating the universe about an axis (O) means the rotation of all points of the universe about the axis. When star S$\rb$ is considered, we are treating it as an extended object. So the rotation must be such that the distance of any point of S$>$ from the axis remains the same throughout (not just the center of mass of the star). Consequently S$>$ will become $<$S after half a period.

\section{Physical effects to detect absolute state of rotational motion}\label{sec4}

What is meant by BR and UR? In the Newtonian conception, all motion is with respect to the absolute frame of reference, and hence BR means that the bucket is rotating with respect to the absolute frame but the universe is static. In contrast, UR implies that the universe is rotating with respect to the absolute frame around the same axis of rotation, but the bucket is at rest. In the Machian interpretation there is no absolute frame of reference, and hence the meanings of BR and UR are more subtle. Consider an observer rigidly attached to the bucket. By looking at the motion of the distant stars, he/she will conclude that either the bucket is rotating and the universe is static or vice versa (similar to what a traveler experiences sitting in a moving train). So the observer can describe his/her observation from two different perspectives, BR and UR, and according to Mach both observations are equally true. An observer rigidly attached with the universe also would have the same freedom in explaining the observed facts. 

In Sec.~I B of Ref.~\onlinecite{ab:5} several interesting physical effects are proposed to distinguish Machian relativity from the absolute nature of space-time. Our comments on those arguments are as follows: 

\subsection{Rotationally induced electromagnetic fields}
HN cited the well-known example of a rotationally induced electric field that Feynman invoked against relativity of rotation. ~\onlinecite{ab:6} A magnetic field is produced when an electrically charged liquid is kept in a rotating bucket. An electric field will be induced in a conducting wire that is placed above the liquid, perpendicular to the axis of the bucket, extending from the axis to the surface of the bucket and tied to the bucket so that it rotates with the bucket. HN questioned whether the electric and magnetic fields can be generated in the UR scenario. To an observer fixed on the bucket, the charged liquid is at rest and hence there should not be any induced magnetic or electric field, which contradicts the conclusion of the observer in the fixed universe. This argument led Feynman to conclude there is no relativity of rotation. 

In Feynman's example the magnetic field (or the electric field in the conducting wire) is produced when the charged liquid rotates with respect to the rest of the universe. If the charged liquid is allowed to rotate in an otherwise empty universe, there will be no magnetic field according to Mach (it is impossible to disprove this assertion with our present knowledge). The magnetic field caused by a rotating charged liquid is calculated with respect to an inertial observer. In the Machian interpretation those who are non-accelerating with respect to distant stars are inertial observers. In both the UR and BR cases, the liquid is rotating with respect to the observer tied to the universe. Hence the same derivation is applicable for the UR and BR cases.

\subsection{A point electric charge on the bucket}

HN pointed out that a point charge located on the rim of a bucket will radiate in the BR case, but it is not clear that similar phenomenon would also occur in the UR case, or at least Mach did not demonstrate that. They further argued that work must be done to rotate the charged bucket at a constant angular speed, and if external energy equal to the energy radiated is not supplied to the bucket, the angular speed of the bucket will decrease. If no external energy is supplied, the loss of kinetic energy of the bucket would equal the energy radiated. In the UR case the angular speed of the universe has to decrease by a magnitude same to that of the bucket in the BR case, but then the loss of kinetic energy in the UR case would be much larger (due to larger mass of the universe) and hence there is a clear distinction between the two cases. 

In the derivation of the radiating energy (Larmor expression), an absolute frame is not essential, and the only consideration is that of acceleration. However, the calculation of the radiation has to be done with respect to an inertial observer (otherwise, a static charge would radiate when an observer is linearly accelerated toward it). With respect to universe-bound observers, the angular velocity of the charged particle is the same, whether it is BR or UR, and hence the derivation and the expression for radiating energy is also identical in both cases. 

In the BR case the emission of radiation takes place at the expense of the kinetic energy of the bucket, and hence the angular speed of the bucket relative to the rest of the universe decreases. In the UR case the bucket tries to rotate with the universe due to inertia. Thus to keep the bucket fixed work has to be done on the bucket resulting in a gain of potential energy of the bucket, and the magnitude of the potential energy would be the same as its kinetic energy in the BR case. As the bucket emits radiation, its potential energy would be reduced by an amount equal to the energy radiated. Consequently the bucket would start rotating; that is, the relative angular velocity between the universe and the bucket would decrease and be the same as in the BR case after radiation. The emitted radiation has a negligible influence on the energy of the universe in both cases. 

\subsection{The Sagnac effect}
The Sagnac effect can be described as the difference in round trip travel time of co-rotating and counter rotating light rays from the standpoint of an observer on the rotating platform. This time difference leads to a shift of the interference fringes, which occurs when two oppositely rotating beams are superimposed on each other in an interferometer placed on the turntable. HN proposed that by performing a Sagnac experiment an observer on the fixed star (of the universe) could determine whether the universe is rotating or not. This proposal demands special attention because it is obvious that no fringe shift would be observed in such an experiment. 

As we have argued there is no simultaneous rotation and revolution of the star in the UR case; only the star would revolve around the axis passing through the center of the bucket. And even for this rotation there would be no Sagnac phase shift. The reasons are as follows. Though several competeting explanations of the Sagnac effect have been suggested from the standpoint of an observer on the rotating platform \cite{ab:14}, for the stationary (inertial) observer, however, the effect has an umambiguous explanation  -- the round trip distance traveled by the co-rotating observer is greater than that of the counter rotating light rays. According to Mach, an observer, who is at rest with respect to a distant fixed star, is an inertial observer for both the BR and the UR cases. So an observer on a distant fixed star would not notice a fringe shift even when the universe is rotating; in contrast, an observer on the bucket would find a change of fringe shift in either the BR or UR case. An important implication of this argument is that the Sagnac effect is caused by the rest of the universe. Hence if a Sagnac-like experiment is conducted in otherwise empty space, there would be no Sagnac effect. 

\subsection{Superluminal velocity}
HN argued that in the BR case, the tangential velocity of an observer on the rim of the bucket of radius $R$ is $\omega R$, where $\omega$ is the angular velocity of the bucket; in contrast, for the UR case the tangential velocity $v$ at a distance $r$ from the axis of rotation would be $\omega r$. Because $r$ could be very large for a distant star, $v$ could be greater than $c$; that is, the tangential velocity of the star could be superluminal. Consequently, the star's motion may produce phenomena like a shock wave at an angle $\sin^{-1}(c/v)$ with respect to $v$. 

In the BR case, the tangential velocity of the observer on the rotating bucket as given in Ref.~\onlinecite{ab:5} is with respect to an observer rigidly attached to the universe. In the UR case, however, the tangential velocity of the distant star is given with respect to an observer on the fixed bucket. Hence the comparison is not proper. Even when the bucket is rotating, the tangential velocity of the distant star with respect to an observer on the bucket could be superluminal. 

\subsection{Bucket water interactions}
HN raised several other objections to the equivalence of the BR and UR cases. One such objection is that the bucket and water interactions would be different in the two cases. For instance, according to Ref.~\onlinecite{ab:5} the time variation of the concavity of the water surface would be different. For a thin vertically shaped bucket in the BR case the maximum concavity occurs soon after the rotation, whereas the concavity should form at the same rate in the UR case for all buckets. HN further stated that in the BR case water droplets fly off tangentially at a constant velocity when the bucket leaks, whereas they would fly off radially in the UR case.

In the UR case the rotating bucket experiment may be interpreted as follows. Initially the water would begin to move with the universe and hence remain at rest relative to the observer. Due to the influence of the bucket, the water starts to acquire the state of motion of the bucket and thus concavity forms. If the bucket is a thin vertical shaped box, the water would acquire the state of motion of the bucket very rapidly, identical to the BR case. When the bucket leaks in the UR case, the motion of a water droplet would be identical to BR to the same observer. The physical interpretation of the motion of the ejected droplet in the UR case could be as follows. An external force (torque) is required in the UR case to keep the bucket fixed (similar to what is needed to rotate the bucket in the BR case) resulting a gain of the potential energy of the system. Due to this potential energy the droplet would move with the linear velocity $v= \omega r$) ($r$ is the distance of the ejection point from the rotational axis) on ejection. After ejection the droplet would acquire the state of the motion of the universe, and hence it will spiral away from an observer tied with the bucket (the universe is rotating and the bucket is fixed). 

\subsection{Multiple buckets}
In Sec.~IIE of Ref.~\onlinecite{ab:5} the issue of multiple buckets was raised. It was argued that Machian relativity is problematic in describing the situation of two coaxial buckets rotating with equal but opposite angular velocities. It is clear that the concavity of the water surfaces in the two buckets would be the same. According to Ref.~\onlinecite{ab:5} Newtonian reasoning can explain this fact by treating the buckets equally and independently, whereas Mach's idea leads to a situation that the universe is rotating in two opposite directions or to a complex situation that one bucket is revolving around the other and that the bucket is rotating around the universe. One may check that if we consider a similar scenario for linear unaccelerated motion (two particles are moving with equal but opposite linear velocities with respect to an observer), similar situation (the stated observer is moving in two opposite directions) would occur. Such situations occur due to introduction of a third independent frame (the universe and the observer act as the third frame in the mentioned examples respectively).      

\section{Heliocentric versus Geocentric Systems}\label{sec5}
A corollary of the relativity of rotation is that the rotation of Earth around the Sun is equivalent to the rotation of Sun around the Earth (as long as no valid argument is raised against Mach's idea, there would not be any problems regarding this equivalence). We confine our discussion to the above corollary without going into the Ptolemaic/Brahean system because our concern is only the relativity of rotation, and because it is not proven that the equivalence of the Brahean and Copernican systems is a necessary corollary of the relativity of rotation, as pointed out in Ref.~\onlinecite{ab:5} (Mach was in favor of equivalence of the Ptolemaic and Copernican systems, but we agree with Ref.~\onlinecite{ab:5} that the observation of the full disk of Venus by Galileo falsifies the Ptolamaic system whereas Brahe's system accommodates such observations). 

In the heliocentric system (the Sun is fixed with respect to the distant stars) the Earth has two kinds of rotation, one about its own axis with a period of 24 hours and another about the center of mass of the solar system with a period of 365 days. Relativity of rotation implies that in the geocentric system the universe is (anti) rotating about the axis of the Earth with a period of 24 hours while the center of mass of the solar system is rotating around the Earth with a period of 365 days. In the later case the distant stars also exhibit motion in such a way that equivalence of the heliocentric and geocentric systems is maintained. (For instance, the center of mass of the solar system would remain static with respect to the distant stars and hence there is no violation of conservation of momentum.) In the geocentric system the planets are also orbiting the Earth, but in a complicated manner. (A common belief is that the planets orbit the Earth rather than the Sun was proven false by Galileo when he discovered the phases of Venus. We would say that Galileo discovered the frame from which the description of the planetary system is simplest and hence most convenient.)

HN raised the question that if heliocentric and geocentric systems are equally possible, then what about luno- or jovo- or venocentric systems? In other words how can simultaneous observations from different planets be accommodated? We say simultaneous observations from different planets are not a problem. We are accustomed to such situations for non-accelerated linear motion. Results of one frame can be transferred to another frame by appropriate transformation equations. 

\section{Discussion}\label{sec6}
It is well known that Mach's idea played an important role in the development of general relativity, although later it was found that general relativity is not a perfect Machian theory. There are cosmological solutions in general relativity that exhibit intrinsic rotation of matter with respect to a local inertial frame.\cite{ab:15} To explain recent observations in the framework of general relativity, exotic/unusual kinds of matter/energy such as dark matter or dark energy have to be invoked, which indicates that general relativity may not be the ultimate theory of gravitation but might be an effective theory. Failure to reconcile general relativity with quantum mechanics despite huge efforts over the years also support such a view. 

We conclude that the recent objections to Mach's idea of the relativity of rotational motion are not correct, and Mach's views of the relativity of rotational motion continue as a viable way of looking at space-time. 

\begin{acknowledgments}
The authors would like to thank Professors Nissim-Sabat and Hartman for cordial and extended communications on the issues involved and for providing useful suggestions. The authors are grateful to Editors of AJP for the help in improving the readibility of the article.
\end{acknowledgments}

\end{document}